# The Geometry of Morphogenesis and the Morphogenetic Field Concept

Nadya Morozova[1], Mikhail Shubin[2]

[1]CNRS FRE 3377, Laboratoire Epigenetique et Cancer, CEA Saclay, France, morozova@vjf.cnrs.fr
[2]Department of Mathematics, Northeastern University, Boston, MA, USA, M.Shubin@neu.edu

This paper presents a set of ideas concerning the connection between biological information, encoded in the cells, and the realization of the geometrical form of a developing organism. Some suggestions are made for mathematical formalization of this connection. The active discussion of this and similar questions at the IHES Workshop (2010) indicated a strong interest in the subject. Therefore, we believe that, although this work is still far from being finalized, it is worthwhile publishing this paper in order to stimulate further discussion. Hence, we strongly encourage the reader to independently consider the ideas, concepts and statements presented here in the framework of an integrated model, as it is possible that some of them may turn out to be unviable, while others may be of some interest and importance.

## Abstract

The process of morphogenesis, which can be defined as an evolution of the form of an organism, is one of the most intriguing mysteries in the life sciences. The discovery and description of the spatial–temporal distribution of the gene expression pattern during morphogenesis, together with its key regulators, is one of the main recent achievements in developmental biology. Nevertheless, gene expression patterns cannot explain the development of the precise geometry of an organism and its parts in space. Here, we suggest a set of postulates and possible approaches for discovering the correspondence between molecular biological information and its realization in a given geometry of an organism in space–time.

First, we suggest that the geometry of the organism and its parts is coded by a molecular code located on the cell surfaces in such a way that, with each cell, there can be associated a corresponding matrix, containing this code. As a particular model, we propose coding by several types of oligosaccharide residues of glycoconjugates.

Second, we provide a notion of *cell event*, and suggest a description of development as a tree of cell events, where by *cell event* we understand the changing of *cell state*, e.g. the processes of cell division, cell growth, death, cell shifting (movement) or cell differentiation.

Next we suggest describing *cell motion laws* using the notion of a "morphogenetic field", meaning an object in an "event space" over a "cell space", which governs the transformation of the coded biological information into an instructive signal for a cell event for a given cell, depending on the position of the cell in the developing embryo. The matrix on a cell surface will be changed after each cell event according to the rule(s) dictated by the morphogenetic field of an organism.

Finally, we provide some ideas about the connection between the morphogenetic code on the cell surface, cell motion law(s), and the geometry of an embryo.

## 1. Biological problem and main postulates for its formalization.

### 1.1. Biological background

All studies on the mathematical/theoretical formalization of morphogenesis can be divided into two main categories: some based on a principle of self-organization and others based on a deterministic concept. We consider the concept of predetermination of a geometrical shape/form of living species to be the most appropriate, and to support the need to build the formalization on this basis, we cite here some important experimental observations.

1. Transplantation experiments in Amphibians (from a donor organism to an acceptor) and in Drosophila (from one part of an organism to another part of the same organism) show the following rule: the specification of the *positional information* is determined by the position of the cell in the organism, and the determination of the *specific cell type* is determined by the specific genomic information of the given cell. (*Positional information* is the information on the location of cell(s) in space related to given reference points inside a developing organism). For example, transplants from a presumptive stomach of one organism (a triton) transferred to the region of the presumptive mouth in another organism (a frog), will develop into a mouth, but the form (geometry) of this mouth will be that of the donor organism (triton) (Spemann, 1938).

2. There are two types of regeneration: morphallaxis and epimorphosis. During morphallaxis the absent parts of the regenerating organism are developed from the existing cells by the changing of their specification and a redistribution in the regenerating organism. An example of this type of regeneration is a redistribution of the cells of the body column of a hydra, forming a new head with tentacles after losing the original head. During epimorphic regeneration, the first step for the cells in the boundary area is dedifferentiation – the formation of the regeneration blastema by rapid cell proliferation; the cells in the blastema become totipotent (i.e., they are able to differentiate into many different cell types) and then start to form a proper organ (e.g. a limb or tail) according to a plan of the whole organism. (Gilbert, 2000).

In both cases of regeneration, during this process cells "move" (change their internal state or location) in order to reach a predetermined morphological and functional structure of the given organism.

3. A cell from a 4–16 cell embryo will behave (grow, divide, differentiate, etc.) differently in a case of further development as a part of a whole embryo, versus a case when it is isolated at this stage and cultivated in cell culture. However, in both cases the final result of development will be the same –the predetermined species-specific form of the given embryo (or larva)(Driesch, 1892). In addition, at the early stages, changing, during the course of an experiment, the position of cells inside the embryo does not prevent normal embryo formation, although initially these cells have been committed to different developmental pathways (Horstadius, 1939).

Thus, the deterministic concept will be the main basis of our formalization of the geometry of development. Interestingly, R. Thom, working on the mathematical formalization of a fundamental law(s) underlying morphogenesis, based on catastrophe theory, ultimately came to the conclusion that it is impossible to create a formalization of morphogenesis that is not based on a "deterministic concept" (Thom,1983,1989)

**1.2. Morphogenetic field notion**

Next, we need to elucidate the notion of "morphogenetic field". The term *morphogenetic field* was suggested by embryologists decades ago to describe the morphology of the developing embryo (Weiss,1939; Waddington, 1956). According to this biological notion, a morphogenetic field can be defined as a group of cells, the location and the future fate of which have the specification within the same boundary (Wolpert, 1977). For example, a zone in the early embryo, including all cells that can potentially participate in the formation of the limb, was named "the field of the limb". Such a field acquires the ability for internal regulation in the case of loss or addition of its parts. In experiments on *Amblystoma* embryogenesis it was shown that each half of the limb disc can regenerate the whole limb at the new position of its transplantation. (Harrison, 1918) The same phenomena of full regeneration of the whole limb from each of several segments of limb disc, divided by impenetrable padding, was shown in experimental and in native conditions. In nature, the phenomenon of frogs and salamanders with multiple legs has been observed as the result of limb disc invasion by the eggs of a parasite trematode, thus cutting the limb field into several parts. This means that the limb field can be considered as an equipotential system, i.e. all cells in this system are committed to form any part of the structure corresponding to this field. (Gilbert, 1991). More recently, the term morphogenetic field has been used, with the same meaning, in a number of works on embryology (Davidson, 1993, 1998, 2002; Bolouri, 2002).

In addition, the notion of morphogenetic field for the mathematical formalization of developmental processes was suggested in some works by R.Thom (1983, 1989), but in a very general manner, with no connection to real biological information.

Despite the differences between biological and mathematical concepts of morphogenetic field, they both reflect two very important observations, namely, that the developmental behavior of a cell depends on the instructive signals from the surrounding space (area), and that different areas in a developing embryo contain precise instructions about the shape of corresponding organs. Because these features cannot be ignored in any model aiming to formalize the developmental process, we continue to exploit the morphogenetic field term in the framework of our model, as a possible convenient tool to describe the connection between biological information, encoded in the cells, and the realization of the geometrical form of a developing organism.

We will use the notion of morphogenetic field in a mathematical sense, meaning by "field" a structure containing a space–time dependent mechanism, which mediates the transformation of biological information, contained in cells, into the corresponding geometrical form of an organism in space–time; or, more precisely, into an instructive signal for a cell motion (cell event), depending on the position of this cell in the developing embryo. In a way, this is an elaboration of the existing (and very general) notion, to give a more concrete meaning.

When introducing mathematical formalization of biological laws, it is important to note that in biology the main law(s) are represented as law(s) based on *coding of biological information* (e.g. protein sequence is coded in DNA). Thus, if the mathematical description of morphogenesis can be made in the framework of field theory, then it should be modified so as to consider the behavior of the objects, whose nature and evolution depends on law(s) of coding of the information (i.e. geometrical information) rather than on energy minimization laws.

### 1.3. Main conjectures for mathematical formalization

We will start formalization with the following **conjectures**:

1. During embryogenesis, each cell undergoes cell divisions, growth, movements (shifts) and expression of specific molecules according to a **determinate plan**, invariant for each living species.

2. This **determinate plan** for development of an organism can be considered as a tree of *cell events* from the initial state of the first cell (zygote) to a final predetermined state of an organism, where under *cell event* we understand "developmental events", such as cell divisions, cell growth (death), cell differentiation and cell shifts.

3. This **determinate plan** is coded by a set of specific biological markers, which, most likely, may exist and be transmitted as a set of cell membrane markers. Our main assumption is that such a code may be provided by a pattern of short oligosaccharide residues of glycoproteins (glycoconjugates) on a cell surface, changing in time and space. It is possible that some other cell surface markers, e.g. specific proteins, may play this coding role; however, short oligosaccharide residues of glycoconjugates have several specific features which make them the most plausible substances for such coding.

4. The general laws for cell events (*cell motion laws)*, namely, the dependence of cell events on coded and positional biological information, have to be the same for all living species, leading to different forms and shapes resulting from different sets of species-specific molecular parameters. Our main goal is to describe these *cell motion laws.*

5. Cell motion laws can be mathematically formulated using the notion of a morphogenetic field.

It is important to note that, in the framework of our model, the cascades of specific molecular events that correlate with pattern formation (e.g. differential gene expression, directed protein traffic, etc.) appear not to be the reason for a cell event. Rather, these cascades are, along with the cell event itself, associated with the "coding information" on a cell surface, or, using another terminology, are realized due to an instruction for the cell from the morphogenetic field of an organism. The concrete signal transduction pathways connecting the "coding information" on a cell surface and the expression of the given sets of genes need to be elucidated.

## 2. Morphogenetic model
### 2.1. Sets of morphogenetic markers on the cell surface can be written in the form of a matrix

We assume that the information regarding the geometry of an organism is contained on the cell surface, in the form of a code composed of biological molecules of a special type. Our prevailing assumption is that, most likely, such a code can consist of oligosaccharide residues of glycoconjugates on the cell surface. There are 12 types of monosaccharide that exist in oligosaccharide residues of glycoconjugates (with six of them being of hexose type), and there have been numerous observations indicating that these oligosaccharide residues are connected with the determination of cellular morphogenetic pathways. (Zablackis et al.,1996; Fry et al.,1993; McNeil et al.,1984; Johnson et al.,1991; Fry 1994; Mohnen, Hahn., 1993 ; Albersheim et al., 1992 ; Riou et al., 1986; Taatjes and Roth, 1991**;** Tran Thanh Van et al., 1985 ; de Vries et al.,1988; Bourrillon and Aubery, 1989 ;Ito et al.,1985; Nemanic et al.,1983 ; Morozova et al.,2006; Ponder,1983 ; Zuzack and Tasca, 1985; Suprasert et al.,1999; Crossin and Edelman,1992; Friedlander et al., 1988;Tan et al.,1987 Edelman,1988). We will use this "oligosaccharide code" assumption in our examples below; however, it is clear that this is not the only possible "cell surface code", and the real biological code (if it exists) should be elucidated by additional joint theoretical and experimental research.

Our next assumption is that this information can be written in the form of a matrix $A_{nm}$, in which every element corresponds to the level of a certain type of oligosaccharide (or monosaccharide) in a given region (section) of the cell surface.

For an easy illustration of this idea, we may consider two sections on the cell surface (left and right), and six types of monosaccharide to be used in a coding, and in this case we will obtain a $2 \times 6$ matrix associated with each cell (Fig.1).

The vector corresponding to each section (i.e. an order, in which we will place the numbers corresponding to the quantity of each type of monosaccharide residue in a section) will have the following structure:

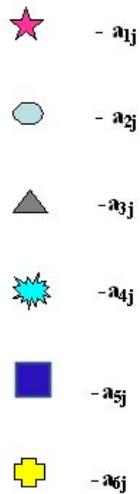

where each symbol defines one type of "coding" residue (mannose, glucose, galactose, rhamnose, fucose or xylose).

When a cell undergoes division, one part of each of the two newly generated matrices, corresponding to two new cells, will be identical to the parent matrix, while the other part will be generated de novo, i.e. filled with new numbers (Fig.1).

We next suggest that filling in the daughter matrices with the new $A_{nm}$ elements after/during each cell division occurs according to a rule $R$, the general principle of which may be the same for all living organisms:

$$R(A^i_{nm}) \rightarrow \{A^j_{nm}, A^k_{nm}\} \qquad (1)$$

where the matrices $A^j_{nm}, A^k_{nm}$ of two new daughter cells are created by application of the rule $R$ to the matrix of the mother cell $A^i_{nm}$.

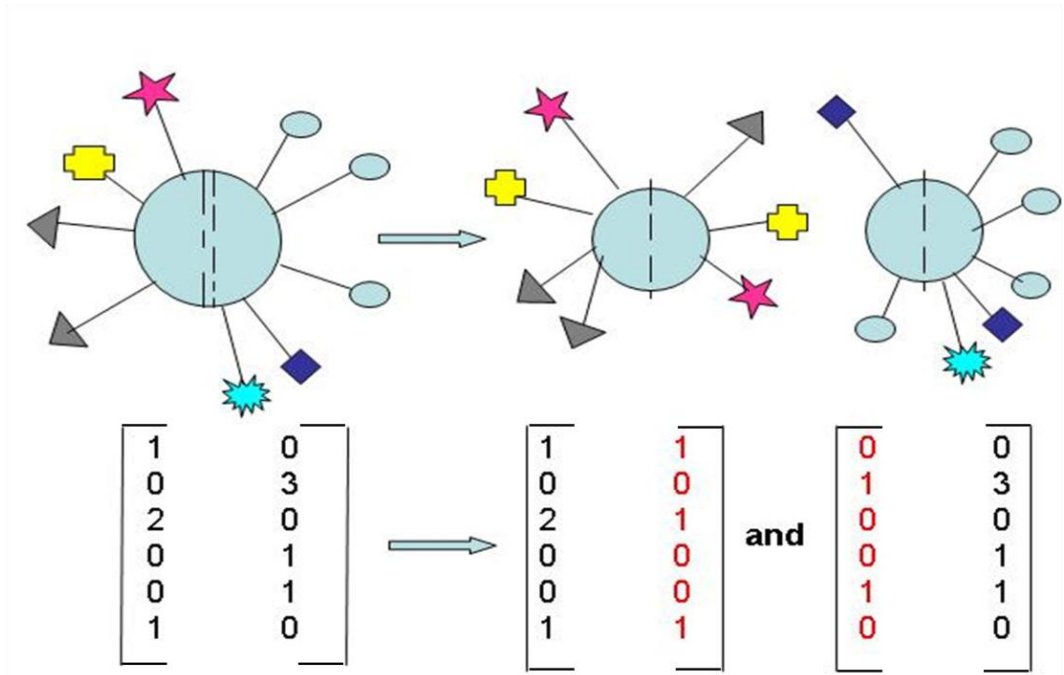

**Fig.1. A simplified two-section example of the matrices corresponding to a mother cell and two daughter cells following cell division. One column of each matrix corresponding to the daughter cells is equal to the mother cell (columns in black), and each new column in the daughter cell matrix is created according to a rule $R_d$ (columns in red).**

In biological terms this means that there exists a rule $(R)$, according to which new glycoconjugate residues will appear on the cell surface in a specific quantities and in a specific locations.

It is clear that a *set of "conservative" elements* $\{A_{nm}\}$ in new daughter cells is determined by the position of the plane of cell division, dividing cell membranes (and thus all sets of its markers) into 2 parts. We can illustrate it with a more realistic example, considering 8 sections on the cell surface with respect to 3 spatial axes, X, Y and Z, and 6 types of monosaccharide to be used in a coding. Then for this case we will obtain a $6 \times 8$ matrix associated with each cell. (Here X, Y, Z axes represent the anterior–posterior (AP), dorsal–ventral (DV) and left–right (LR) axes, normally used in embryology).

If a cell divides symmetrically with respect to the membrane surface, and only in planes X=0, Y=0, or Z=0 with respect to the coordinate system defined for the zygote, then during each cell division one set of 4 columns from the 8-column matrix $A_{nm}^i$ of the mother cell $i$ will be equal to the corresponding set of matrix $A_{nm}^j$ of a daughter cell $j$, and the other set of 4 columns of $A_{nm}^i$ will be equal to matrix $A_{nm}^k$ of a daughter cell $k$. The remaining 4 columns (of a total of 8) in each new cell $j$ and $k$ will be created *de novo* after each cell division.

For example, if the plane of cell division is X=0, columns 1,2,3 and 4 of the mother cell matrix will be equal to the same columns of the "left daughter cell" and columns 5,6,7 and 8 to the "right daughter cell"; if the plane of cell division is Y=0, columns 1,3,5 and 7 of the mother cell matrix will be equal to the same columns of the left daughter cell and columns 2,4,6 and 8 to the right daughter cell; etc.

Thus the proposed model implies that the determination of the position of the cell division plane is realized by the rule $R_d$, by defining the new matrices of new daughter cells. The elements $A_{nm}^j$, which, according to $R_d$, will be equal to those in the mother cell $A_{nm}^{j-1}$, will thus determine the border between 2 new daughter cells, and thus, the position of the cell division plane.

Generally speaking, a cell division plane can divide a cell into 2 parts not symmetrically related to the membrane surface, and the cell division plane can be under any set of angles α,β,γ to embryo axes X, Y, Z. On the other hand, it is clear that, in the framework of the very simple matrix representation of "cell surface coding" used above, it must be postulated that only those locations of the cell division plane that cut the cell surface at a boundary between two "coding sections" are permitted. Thus, for considering the realistic potentiality of cell division plane orientation, the suggested meaningful quantity of "coding sections" on a cell surface should definitely be greater than 8 (16, 32, or more). If one wishes to avoid such a "boundary postulate", then, instead of a matrix, cell surface information should be written as a 3-dimentional array, with the additional dimension corresponding to the location of each marker inside each "coding section".

Changes in the quantity and composition of glycoconjugate residues on the cell surface during cell growth between its birth and vanishing (when it divides or dies) will occur according to another rule(s) $R_l$. These rules have to account for the processes of both appearance and disappearance of the residues on the cell surface.

We may suggest, therefore, that there is one rule $R_l$ for each type of *cell event $S_l$*.

## 2.2 Cell event definition

We will assume that each cell *i* can be described by a cell state function $\psi^i(t)$,

$$\psi^i(t) = \left(A_{nm}^i(t), \mathbf{D}^i(t)\right)$$

where **D** is a parameter showing the position of cell *i* in the embryo with respect to the zygote. This can be given in several ways, e.g. as a cell lineage and generation number, or, alternatively, as $\mathbf{D} = \sum \mathbf{d_k}$ where $\mathbf{d_k}$ is a vector perpendicular to the cell division plane and **k** is the number of cell divisions that have occurred from the zygote to the given cell (in the case of no other displacements, such as shifts).

In addition, we may propose an alternative possibility to mark the position of cell *i* in the embryo. It makes sense to assume that the cell surface markers on the zygote may have some additional specificity, discriminating them from all other markers of the same type, synthesized later in the course of development. In this case, the position of each cell in the embryo can be marked with respect to zygote membrane/surface, according to these specific markers.

By *cell event* we will understand the changing of the *cell state* $\psi^i(t)$, i.e. as a result of the cell event, cell matrix $A^i_{nm}(t)$, cell position $\mathbf{D}^i$, or both will be changed, generating a new cell state $\psi^i(t_1)$. This change can occur as a result of the processes of cell division, cell growth, cell death, cell movement in space (shift) or cell differentiation.

We can formalize all possible cell events $S_l$ as follows, subdividing them into 3 main types:

$S_1$ = cell division in plane W, with angles α,β,γ to embryo axes X,Y,Z (AP, DV,LR)
$S_2$ = cell growth, characterized by cell growth rate q.
$S_3$ = cell shift, characterized by cell shift vector **h**.

Importantly, $S_2$ corresponds to 3 different processes:

$S_{21}$—cell grows (cell growth rate q >0)
$S_{22}$—cell undergoes apoptosis (programmed cell death) (cell growth rate q<0)
$S_{23}$—cell differentiates (cell growth rate q= 0)

depending on the sign of a cell growth rate q, $q = \frac{dV}{dt} = \oiint a \, dA$, where $a$ is a local growth rate at point x on a cell surface C and $dA$ is an area element on that surface. A cell growth vector $\mathbf{B}(\mathbf{x})$ on the cell surface is then defined as $\mathbf{B}(\mathbf{x}) = a(\mathbf{x})\mathbf{n}$, where **n** is the unit normal to cell surface at point x. However, for the process of apoptosis it might be better to associate an additional type of cell event.

If we present **Cell event as an operator**, acting on a cell state $\psi^i(t) = \left(A^i_{nm}(t), \mathbf{D}^i(t)\right)$ it should be able to give the following types of results for different types of Cell Events:

Division: $\left(A^i_{nm}, \mathbf{D}^i\right) \rightarrow \left(R_1(A^i_{nm}), g_1(\mathbf{D}^i)\right)$ and $\left(R_2(A^i_{nm}), g_2(\mathbf{D}^i)\right)$
Growth: $\left(A^i_{nm}, \mathbf{D}^i(t_1)\right) \rightarrow \left(R(A^i_{nm}), \mathbf{D}^i(t_2)\right)$
Differentiation: $\left(A^i_{nm}, \mathbf{D}^i\right) \rightarrow \left(R(A^i_{nm}), \mathbf{D}^i\right)$
Apoptosis : $\left(A^i_{nm}, \mathbf{D}^i\right) \rightarrow \emptyset$
Shift: $\left(A^i_{nm}, \mathbf{D}^i\right) \rightarrow \left(R(A^i_{nm}), f(\mathbf{D}^i)\right)$

where a function $g(\mathbf{D}^i)$ determines cell positions of the daughter cells $\mathbf{D}^j$ and $\mathbf{D}^k$ of a cell $\mathbf{D}^i$, a function $f(\mathbf{D}^i)$ determines cell position after a shift **h**. The form of functions g and f, as well as the mathematical description of the relation between $\mathbf{D}^i(t_1)$ and $\mathbf{D}^i(t_2)$ for the cell event Growth depend on a choice of a parameter **D** showing the position of cell *i* in the developing embryo, and, in particular, on how a size of a cell and a form of a cell surface will be included in a cell position **D**.

In a very simplified case of development with an assumption that cells divide symmetrically in relation to their membrane surface and with only 3 possible division planes (X=0, Y=0, Z=0), we will have only 3 types of cell division event:

$S_{11}$ —cell division in plane X=0
$S_{12}$—cell division in plane Y=0

$S_{13}$—cell division in plane Z=0

Fig. 2 presents an illustration of this formalization to one step in a 3-cells embryo development between 2 time points, corresponding to the stages $G_a$ and $G_k$ of an embryo, and for this step the cell event for each cell will be:

For cell number 1: $S_{12}$ (cell division in the plane X=0)

For cell number 2: $S_{11}$ (cell division in the plane Y=0)

For cell number 3: $S_{21}$ cell growth with cumulative growth rate q>0 ($q \approx 2$), where the growth vector has direction parallel to the y-axis.

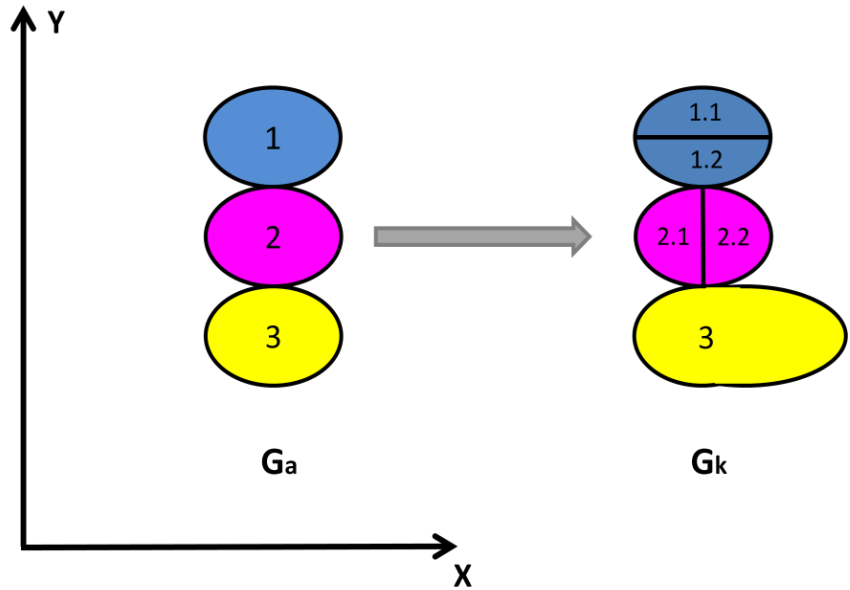

**Fig.2. Illustration of *cell event* concept in 3-cell embryo development.**
**See text for details.**

## 2.3. Evolution of embryo domains' surface geometry depends on cell events within a given domain

The growth and development of an organism in space can be described as the dynamics of the boundaries of the corresponding **domains** of the developing embryo, morphologically distinct from its other parts (e.g. tissue, organ, limb bud or growth point). We will denote as $G_Y(t)$ the surface of the domain Y, which will be a smooth closed surface in $\mathbb{R}^3$, depending on time *t* as a parameter. For any $x$ at $G_Y(t)$, we will denote as an external unit the normal vector to $G_Y$(t) at $x$ by **n**, so that $\mathbf{n} = \mathbf{n}(x, t)$.

Given $G_Y(t)$, we will assume that, at the next moment, $t + dt$, the form of the surface $G_Y(t + dt)$ is such that it consists of all points $x + \sigma(x,t)\mathbf{n}(x,t)dt$, where $\sigma(x,t)$ is a scalar function on $G_Y(t)$ at every given moment of time $t$.

To avoid a gauge transformations indeterminacy (existence of many equivalent presentations of the same family of surfaces which differ by choice of coordinates), we may present a growing embryo domain surface in the form

$$G_Y(t) = \{ x : u(x) = t \}$$

where $u(x)$ is a real-valued function on $\mathbb{R}^3$ with sufficiently regular level sets $S(t)$. In this case the distance between infinitesimally close surfaces $G_Y(t + dt)$ and $G_Y(t)$ at point $x$ is $\sigma(x,t)dt$ (by definition of σ) on the one hand, and using the geometric meaning of the gradient, it is $|\nabla u(x)|^{-1} dt$ on the other hand. This leads to the equation for the evolution of an embryo domain's surface:

$$\sigma(x,t)|\nabla u(x)|^{-1} = 1 \qquad \text{provided } u(x) = t, \qquad (2)$$

which may be written as a system of equations.

We can then see the interconnection between the shape of the domain Y, the form of its surface $G_Y(t)$ and the set $\{S_l\}_{G_Y}$ of all cell events that have occurred inside the domain Y enclosed by surface $G_Y$ during time period dt (Fig.3):

$$\sigma(x,t) = \sigma(\{S_l\}_{G_Y})$$

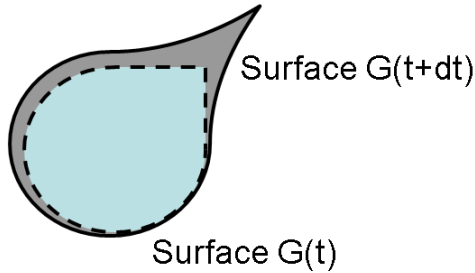

**Fig.3 The shape of a morphological domain Y (= the form of its surface $G_Y$) depends on a set of all cell events that have occurred inside the domain Y enclosed by surface $G_Y$ during time period dt.**

Thus, knowing all cell events in the domain Y during the time period *dt*, we would be able to calculate $\sigma(x,t)$, and next, given $G_Y(t)$ at moment *t* and $\sigma(x,t)$, we can find $G_Y(t + dt)$. Correspondingly, the volume $V_{G_Y}$ of the domain Y may be calculated as:

$$\frac{dV_{G_Y}}{dt} = \oiint_{G_Y} \sigma(x,t) \, dG$$

Cases of branching of a morphological domain into 2 new domains or the appearance of a separated domain inside an existing one are more difficult and require additional conditions for defining boundaries between domains to be included.

## 3. Formalization using the notion of morphogenetic field

Next we will discuss the connection between the morphogenetic code on a cell surface, cell events and the geometry of an embryo using the notion of a morphogenetic field (MF).

We will mean by the term "morphogenetic field" an object governing a law (or a set of laws) for transforming coded morphogenetic information into an instructive signal for a cell event $S_l$ for a given cell, depending on the position of this cell in the developing embryo.

We may consider the existence of a hierarchical structure between the morphogenetic field of the whole organism and the local morphogenetic fields of the organ, the tissue, the morphologically distinct domain of the developing embryo, and of a single cell, assuming that the morphogenetic field of each tissue/domain in the developing embryo is a function of the morphogenetic fields of its cells, the MF of each organ is a function of the MFs of its tissues, and the MF of the whole organism at each stage is a function of the MFs of its organs (or morphologically distinct domains during early embryogenesis). The sense/scope of the instructive signals in each type of the MF in this hierarchy (i.e. MF of organ, MF of tissue, etc) should be different. Possibly, for MFs of higher levels in the hierarchy than the initial "morphological domain", the notions of *tissue event* and *organ event* should be introduced, indicating the signal for new tissue/organ formation. For example, the formation of each somite could be considered as a tissue event, and limb regeneration - as an organ event.

We will start our proposition with the description of the MF of a morphologically distinct domain of the developing embryo, as the initial step in this hierarchy.

### 3.1. Morphogenetic field can be defined as a function of a set of morphogenic markers

We conjecture that the morphogenetic field of any morphological domain of the developing embryo at a given moment in time is a function of the morphogenetic information of all cells located in that domain.

Thus, a mechanism influencing the evolution of the domain in space–time depends on (is a function of) the whole set of all matrices, corresponding to all cells belonging to the given domain, together with a factor of interactions between neighbor cells and a factor depending on the external media. Let us denote a function corresponding to MF influence as $M_F$. Thus, we can write:

$$M_F(t) = M_F\big(\{A_{nm}^i(t)\}, M_{int}(t), M_m(t)\big) \qquad (3)$$

where $\{A^i_{nm}(t)|i \in Y\}$ is a set of matrices for all cells $i$ within embryo domain Y, $M_{int}(t)$ is a factor of interactions between neighbor cells, influencing $M_F$ formation, and $M_m$ is a factor depending on the external (media) conditions.

Anyhow, it is possible that a set of cell surface markers will be changed depending on the external conditions of a cell. In this case, information regarding the effects of $M_{int}(t)$ and $M_m(t)$ on $M_F$ formation can be included in $\{A^i_{nm}(t)\}$, and we can write the simplified version of (1):

$$M_F(t) = M_F(\{A^i_{nm}(t)\}) . \tag{3'}$$

Next we postulate an existence of non-linear operators $R_l$ of MF, corresponding to each cell event $S_l$, such that filling in the new matrices with the new elements after/during each cell event $S_l$ occurs according to operator $R_l$. For example, for event $S_1$ (cell division) it will be operator $R_d$

$$R_d(A^i_{nm}) \to \{A^j_{nm}, A^k_{nm}\}$$

where $A^j_{nm}, A^k_{nm}$ corresponds to 2 new matrices of 2 new daughter cells after the division of the cell $A^i_{nm}$. For event $S_2$ (cell growth) it will be operator $R_g$

$$R_g(A^i_{nm}(t_1)) \to A^i_{nm}(t_2)$$

where $A^i_{nm}(t_1)$ corresponds to the matrix of cell $i$ just after cell division, and $A^i_{nm}(t_2)$ corresponds to the matrix of the same cell $i$ after a period of growth. Thus, $R_l$ is a rule for creating new matrices depending on the matrices of the previous state of the cell.

### 3.2 Morphogenetic field defines cell event for each cell

The next, very important, idea in our model is that, having a rule $R_l$ for filling in the new matrices depending on the matrices of the previous state of the cell, we must consider the influence of MF and cell position **D** in the embryo for this process.

$$S_l = S_l(M_F, \mathbf{D})$$

This means that a cell with the given set of coding molecules on its surface will undergo different cell events and thus will have different rules $R$ for creating the next matrices for the next cell state of this cell, depending on the position of the cell in the embryo. The mechanism regulating this is provided by function $M_F$ of MF. The best example of such regulation is transplantation experiments in Amphibians (see Biological Background section), where cells of a triton committed for producing a stomach, after transplantation into a new (mouth) location in another organism (frog), develop a mouth. This can be explained as a command by MF of a tissue in a new cell position to implement a new chain of *cell events* for these cells, resulting in the proper structure for this organism. The fact that the form (geometry) of this mouth will be that of the donor organism (triton) means that the appropriate cell events depend both on MF field instructions, and on the content of coded biological information (matrices) of the cells $A^i_{nm}$.

Thus, the morphogenetic field, depending on the set of all information $A_{nm}^i$ of its cells, defines a cell event for each cell depending on its position in the embryo (and, possibly, also on $A_{nm}^i$ of the given cell) and $R_l$, corresponding to each type of cell event, works for filling in the new cell matrices $A_{nm}^i$. In the frame of the model, two biological events – cell motion in MF (cell event) and the changing of the oligosaccharide code (matrix) on the cell surface – are supposed to take place during the same time interval, but in three different steps inside this interval: matrix -> MF action -> cell motion -> new matrix.

*Important comment:*

There is another possible way to present the idea of the $A_{nm}^i$, $R_l$, $S_l$ and MF connection based on a different understanding of the nature of MF and thus, resulting in a different way of describing it. The set $A_{nm}^i$ of oligosaccharides on the cell surface (their quantity and composition) may unambiguously determine $R_l$, which also unambiguously determines the corresponding cell event $S_l$. In this case, MF can be understood as a field of cell matrices $A_{nm}^i$ determined for each point (cell) in $\mathbb{R}^{3+1}$ with a non-linear operator $R_l$ acting in this field. However, this representation cannot explain some of the biological data.

### 3.3. Development of an organism can be presented as a Tree of Cell Events.

It is clear, that the final form of any object (tissue, organ, organism) is **invariant** for each living species (with some small fluctuation about which we will tell later). We may assume that **the MF is capable to decipher the initially coded on the zygote surface information and to transform it into a "coded tree of cell events"** for achieving the final shape of an organism.

Thus, if we formalize the development of an organism as a Tree of Cell Events, then the main **Statement (1)** will be the following (Fig.4):

MF determines the morphogenetic effect $M_F$ applied to a given cell, such that the resulting cell events in the domain containing the cell will cause the minimization of the **distance between the initially coded tree** of cell events for achieving the final shape of this domain and the practically (biochemically) **possible tree of cell events from the current state** at a given time point.

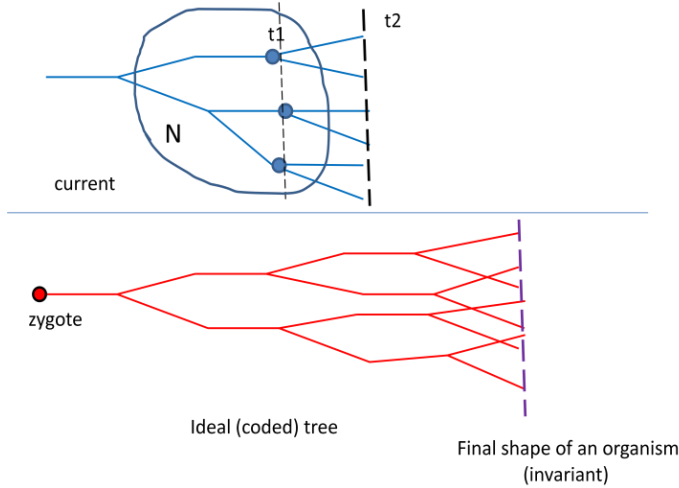

**Fig.4. Example for the concept of the "minimization of the distance between trees".**
**There are 3 cells inside domain N, and each cell has 2 possible (allowed) variants of cell events. From these 6 possible cell events which can be accomplished based on all information in N, will be chosen the ones which will make the real tree more close to the ideal one.**

The real organism attempts to make its MF identical to its corresponding coded MF ($MF'$) by minimizing the discrepancy between them at each time point. Thus, at each time point **t** all cell events leading to the next step (**t**+1) will be dictated by minimization of the discrepancy.

### 3.4. General model

Our main assumption here will be that the MF of a morphological domain (tissue or organ), being a function of coded biological information (a set of matrices of all cells within this domain), also contains a possibility of "measuring" the discrepancy between the current geometrical form of the domain (tissue, organ) in $\mathbb{R}^{3+1}$ and the final geometrical form of the surface of this domain in $\mathbb{R}^{3+1}$. One possibility is that such a measurement is equivalent to the measurement of the discrepancy between the current form of the MF and its final coded (prescribed) form. The details of this possibility we will discuss later (section 4.1). Another possibility may be connected with our previous assumption, that cell surface markers of the zygote may have some additional specificity, thus determining the position of each cell in the embryo with respect to the zygote membrane/surface.

Let us decide that this discrepancy for the domain Y can be measured as a *correspondence function* $F_c$ between surface $G_Y(t)$ at moment *t* and surface $G_k = G_Y(t_{final})$:

$$F_c = F_c(G_k, G_Y)$$

Next we assume that the influence (effect) $M_{F_Y}$ of the morphogenetic field of the domain Y on the cell *i* belonging to this domain depends on the *correspondence function* $F_c$

$$M_{F_Y} = M_{F_Y}\bigl(F_c(G_k, G_Y)\bigr), \tag{4}$$

and results in one of the possible cell events $S_l$, changing the cell state $\psi_i$:

$$M_{F_Y}\psi_i(t_1) = S_l : \psi_i(t_1) \to \psi_i(t_2). \tag{5}$$

Using these notations, we can formulate the following **Statement (2)**:

*MF determines the morphogenetic effect $M_F$ applied to the given cell, such that the resulting cell event will cause the minimization of the correspondence function $F_c$ between the current state of the surface of the domain $G(t)$, containing the given cell, and the final surface of this domain $G_f(t_{final})$* (Fig.5).

On the other hand, using the Statement 1 we can define a *correspondence function $F_c$* not as a discrepancy between surfaces $G_Y(t)$ and $G_k = G_Y(t_{final})$ of the embryo domain, but as the distance between real and "coded" trees of cell events in the sections of MF :

$$F_c = F_c(MF', MF).$$

It is clear, that these two definitions will be equivalent if we are able to embed the tree of cell events in the topological space, reflecting the real topology of an organism. The possibility to do so will be discussed later.

$F_c$ is connected with $M_F$ in the following way: $F_c$ determines the pathway from the current to the final morphological state of the given domain as a tree of potential cell events in $\mathbb{R}^{3+1}$, and $M_F$ determines the local motion law ($S_l$ at the moment *t*) along this way (real cell event in each real time point).

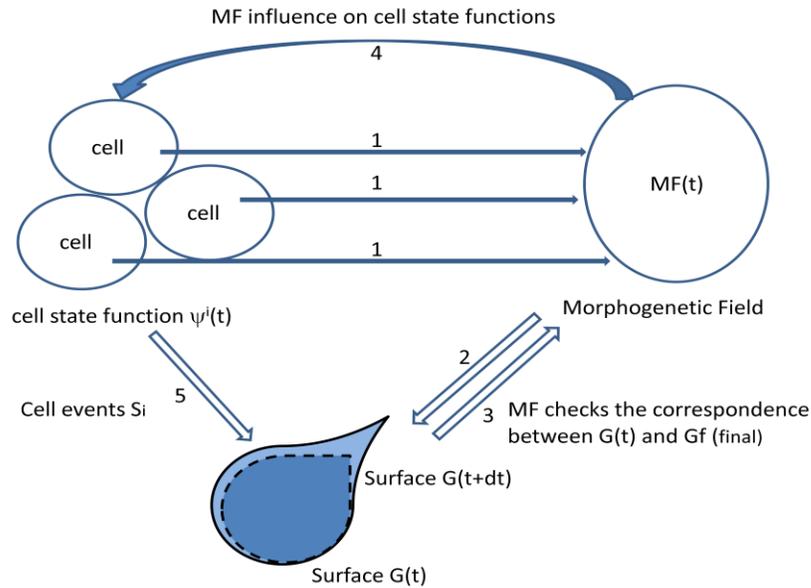

**Fig.5. Schematic representation of the general model**
1.Set of cells determines morphogenetic field (MF)
2, 3. MF evaluates the geometry (position in space) of the surface G(t) (dashed line) and the correspondence between G(t) and $G_f$(final).
4. Influence of MF on the cells ($M_F$).

**5. As the result of MF instruction, each cell from the set undergoes the corresponding cell event for creating a proper new object with a determined surface $G(t + dt)$ (solid line).**

The proposed idea means that, for example, in a situation when "cell 1" of the embryo presented in Fig.2 is destroyed at the stage $G_a$, the cell event for cell 2 will be different – it will be $S_{12}$ (cell division in a plane X=0), thus replacing the cell 1, and the growth rate q for cell 3 will be considerably decreased. Next, all 3 cells of a new 3-cell embryo will undergo 3 corresponding cell events, as described above for the normal 3-cell embryo, thus reaching the proper stage $G_k$.

### 3.5. Biological observations supporting the general model hypothesis

Our general model, to some extent, can be confirmed in its basic form by an important biological observation. The model is based on the hypothesis that glycoconjugates located on a cell surface, concurrently play a role in the coding of the final geometry of an organism and in the signaling for cell events at each moment of time. This proposed dual function of matrices $A_{nm}^i$ can be confirmed by the following biological observation, which may be formulated as the following **Statement (3)**: *the precision of determination of the geometrical form of an organism by biological information is in inverse proportion to the level of totipotency of its cells.*

For example, the animal body is determined unambiguously (e.g. for mammalians: 1 head, 4 paws and 1 tail, and everything at the specified locations) and normally animal cells have a very low level of totipotency (this means that differentiated animal cells cannot undergo de-differentiation followed by re-differentiation into other organs and tissues in cell culture). The forms of plants are not unambiguous, and only the form of specific plant structures, such as flowers or leaves is determined uniquely, but not the final form of the whole plant (for example, in most cases, leaves and branches can appear in random quantities and in multiple possible places). This correlates with the fact that differentiated plant cells have a very high level of totipotency (i.e. one can easily obtain de-differentiation and re-differentiation of plant cells to form a new organism (or any plant organ) in cell culture).

In the framework of our suggested model, we can explain this phenomenon in the following way: in plants the cell surface code may have more redundancy (degeneracy); this influences, at the same time, the non-stringency for the final geometry of an organism, and the flexibility of conformity between a given cell matrix and its corresponding $R$, which allows, accordingly, a wide spectrum of possible meaningful cell events in the frame of the same initially coded information, thus increasing the cells' potency to differentiate into all the specialized tissues of the developing organism. For animals, the cell surface code may have much lower redundancy, and this causes a stricter determination of the final form of the organism, together with stringent individual cell fates for their cells, giving no possibility for alternative cell events.

Another important confirmation of our hypothesis is the behavior of protoplasts in cell culture. Plant cells have a cell wall, which covers the plant cell outside the cell membrane. Due to this fact, in plant cells all oligosaccharides of glycoconjugates of a cell surface are located on the surface of the cell wall, while in animals they are on the surface of the cell membrane. Protoplasts are plant cells without a cell wall, which can be removed from cells in cell culture using a specific enzyme. After cell wall removal, protoplasts produced from terminally differentiated cells of a plant (leaves, fruits, etc.) start to divide, forming at the same time a new cell wall. However, the new cell wall does not have any morphogenetic information, thus all cells continue to divide eternally, producing a callus (an unorganized mass of undifferentiated cells). This means that the morphogenetic information required for proper development was removed with the cell wall, supporting the hypothesis regarding the significance of molecules on the cell surface for coding morphogenetic information.

Cells in a callus can then be induced by different plant hormones to produce different organs (roots, shoots, etc). Important, that plant hormones which exist in an organism in extremely small concentrations, are able to change the whole direction of development of cells in a callus: for example, depending solely on the ratio between two plant hormones, all callus cells will be governed to produce roots or shoots. This can be interpreted as a restoration of part of the morphogenetic information on a cell surface by specific activation by the hormones of some of the molecular signaling pathways responsible for turnover of oligosaccharides on the cell surface; the final pattern of this morphogenetic information (spectrum of oligosaccharides), which determines the subsequent organ development, depends on the interplay between hormones controlling signaling pathways responsible for each oligosaccharide. The importance of plant hormones for glycoconjugates production was already shown experimentally (Tran Thanh Van et al.,1985.;de Vries et al., 1988)

Finally, in the framework of our model, the process of carcinogenesis can be understood as a loss by cancer cells of the ability to accept the correct signals of the MF to which they belong. The explanation for this may be that the mutations that lead to cancer progression influence the pathways responsible for the correct turnover of oligosaccharides on the cell surface, thus breaking from normal cell behavior dictated by MF. Substantial experimental data showing that some cell surface proteins, primarily glycoproteins, are altered in cancer cells, may be considered as additional support for our suggested hypothesis.

## 4. Some ideas on a possible mathematical description of MF
## 4.1 Mathematical description of MF as a section of a fiber bundle

Now we come to the question: what is the nature of the MF field and how can we describe it? We understand by MF **a species-specific object in the space of events (cell events) over the space of cells** (**cell space**), having the capability of decoding biological information and converting it into the corresponding cell event depending on the location of the cell in the space–time of a developing embryo.

We suggest that it may be possible to formalize this idea using a description of a morphogenetic field as a section of a fiber bundle over a base, where the base contains all the cells existing in $\mathbb{R}^{3+1}$ space during the entire life of the embryo, while the possible cell events are considered as a fiber (Fig.6A). Thus, we will introduce

1. a space-time X which consists of points $\tilde{x} = (\psi, t)$ where $\psi$ is a cell state of a cell, defined by its matrix and a position of a cell in the embryo, $t$ is a time; we consider as X all sets of cells/ cell states existing in the embryo's entire developmental history;

2. a fiber bundle over X with the bundle space $E$ and the projection map $p: E \to X$, where the fiber $E_{\tilde{x}}$ over a point $\tilde{x}$ in X consists of all possible cell events $S_l$ for the cell $\tilde{x}$ in a given cell state. This means that each fiber has the following structure (Fig. 6B): a union of three matrix spaces, corresponding to three types of cell events, united by a singular point at $S_1=S_2=S_3=0$, corresponding to the process of differentiation (see section *2.2 Cell event definition*). An operation in this fiber space -a composition of cell events -is not commutative.

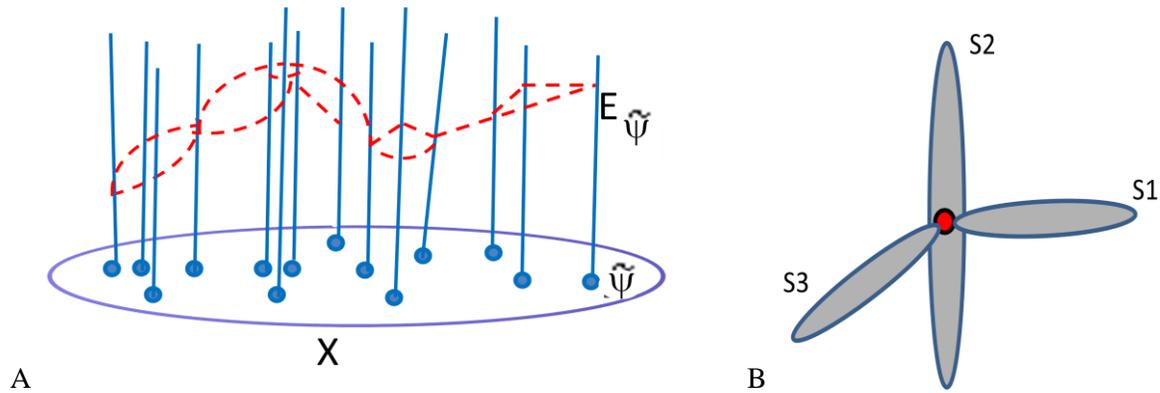

**Fig.6. Schematic representation of a morphogenetic field as a section of a fiber bundle over a base.**
**(A) Base X represents all cells states of all cells during the entire life of the embryo, all possible cell events are considered as a fiber $E_X$. The morphogenetic field is represented as a section of the fiber bundle (in red), choosing the given cell event (point) on each fiber corresponding to each point $\tilde{x}$. (B) A fiber structure. $S_l$ represent three matrix spaces, corresponding to three types of cell events (cell division, cell growth and cell shift, respectively). The singular point in red of their intersection corresponds to cell differentiation.**

In this way, a section s of a fiber bundle

$$M_F(\tilde{x}) \to S_i$$

corresponds to a law for choosing/prescribing a cell event $S_l$ to each cell $x$ in its cell state $\psi^i(t)$, where $M_F$ depends on a set of matrices in the embryo domain, corresponding to the given MF. We may consider a space of the sections of the fiber bundle (MF space). Next we will consider an ordered set of cell events in a time interval, corresponding to section s, as a motion in morphogenetic field, and try to elucidate the law(s) of cell motion. We will assume that function $C$, determining cell motion, is a function on the bundle space $E$ of a point $S_l$ on the fiber over the corresponding point $\tilde{x}$, and $\tilde{x}$ itself:

$$C = C(\tilde{x}, S_l) . \tag{6}$$

We will choose those local sections s of a fiber bundle corresponding to the given MF that will satisfy the equations of motion in MF, which may be found, according to our general model, from the condition:

$$\delta F_c[s] = 0 \tag{7}$$

where the correspondence function $F_c[s]$ will be defined as a real-valued function of the sections of the bundle:

$$F_c[s] = \sum_{\tilde{x} \in X} C(s(\tilde{x})), \tag{8}$$

and the variation $\delta F_c[s]$ is taken with respect to all possible variations of the section s . Here, for discrete parameters, the variation is to be understood as the set of all differences $F_c[s] - F_c[s']$, where for every x ∈ X the (variable) section s takes values that are the closest possible to the corresponding values of $s'$, a section which corresponds to "coded" MF ($MF'$). If the most general form of $C$ were known, then imposing conditions (7) on the equation (8) would in principle result in a set of differential equations, assuming the sum converges to a well-defined integral. These differential equations could be interpreted as the equations of (cell) motion.

Thus, in the framework of the suggested model, a cell event, corresponding to each cell, is dictated by the morphogenetic field by selecting a point on each fiber corresponding to each point on the base. The section of MF of the whole organism consists of numerous sections for all morphological domains and tissues, finally combined in organs. The development of the embryo in time corresponds to a t-dependent 3-dimensional object moving along a 4-dimensional MF section of the fiber bundle.

A cell motion (a path) in MF can be represented as a tree of cell events in this section of fiber bundle (shown on the Fig.4 and Fig.7, upper), i.e., as a graph where each vertex represents a cell event (a point in a section of fiber bundle) corresponding to the *cell state* $\psi$ on the base, while edges connect cell events with causal inference. This will be described in details in sections 4.2 and 4.3.

**4.2 Internal symmetries in the MF structure.**

Next we will look for the internal symmetries in this total space of cell events which may suggest the general form of cell motion laws.

First of all, it is clear, that additional conditions may be imposed on the MF structure and cell motion laws by analysis of characteristics of a natural graph structure on X with two types of edges described as follows:

1. edges connecting neighbor (adjacent) cells on the same time level;
2. edges connecting each cell with its daughter cells.

Cell motion in MF, which is based on the trees of corresponding cell events (shown on the Fig.7, upper graph), is tightly connected with the genealogical graph(s) of cells in the developing organism (Fig.7, lower graph, where each vertex corresponds to a cell in an organism); however these two graphs (or their parts) may be isomorphic only in the cases when a developmental program consists of rapid cell divisions occurring without any other cell events. On the other hand, there are cases when cell motion in MF is completely independent of genealogical graphs, e.g. during a process of regeneration in Hydra (morphallaxis, see section 1.1), where cells undergo a changing of their positions inside an organism accompanied by changing of their morphology, thus creating a new shape without additional cell divisions. In this case a cell motion in MF will be represented as a set of individual trajectories in MF for each cell, consisting of shift, growth and differentiation cell events, integrated in a section(s) due to the corresponding graph structure on X which includes edges between neighbor (adjacent) cells (for different time levels). Thus, each section of MF should be designated with relation to information borne by both types of edges (1 and 2) in the graph structure on X (Fig.7).

Formalization of these graphs structures throughout all types of living organisms may give an idea about some general features of a total space (MF) structure. Together with the assumption that the final form of any object (tissue, organ, organism) can be considered as an **invariant** with respect to the process of species development this may lead to the discovery of **internal symmetry(s)** of the field.

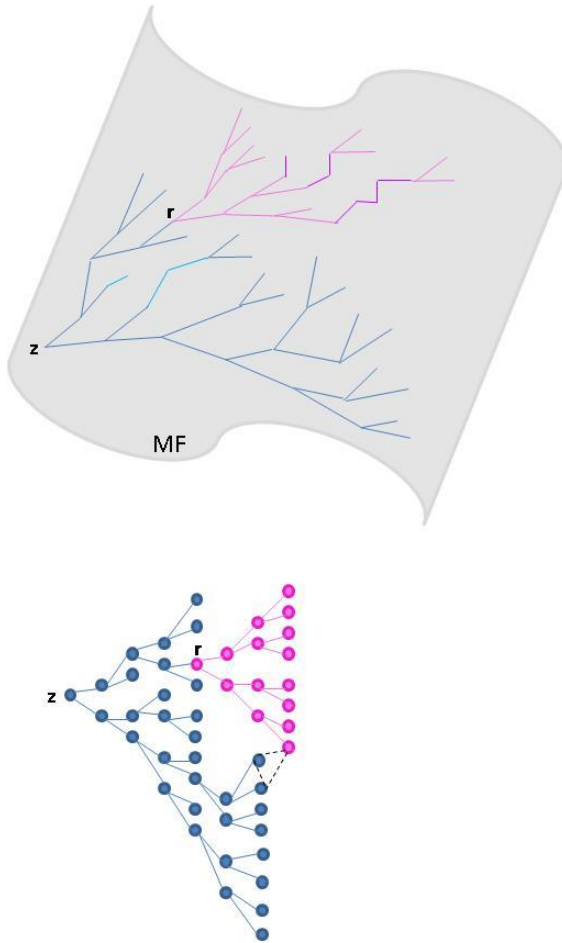

**Fig.7 Tree of Cell Events in Morphogenetic Field.**
**Each vertex on the lower graph corresponds to a cell in an organism; the graph represents a genealogical tree of cells. The triangle in black dotted lines shows an example of edges bearing an information about neighbor cells in an organism.**
**Each vertex on the upper graph represents a cell event, corresponding to *cell state* $\psi$ on the base. Cell motion (a path) for the cell *z* is the whole tree, while cell motion (a path) for the cell *r* is a tree marked in pink color. Outgoing edges from the vertices, which correspond to cell events other than cell division ones (e.g. cell growth, apoptosis, differentiation), are shown in different tints (light-blue or purple-pink).**

## 4.3 Notion of Realization of cell event.

There is a following interconnection between edges in MF (section) and edges of the graph on a base, which still have to be described in mathematical terms.

Each cell in its *cell state* $\psi$ on the base has a prescription for a cell event from the MF, which we mark as a point on a corresponding fiber. The realization of this prescription means a corresponding edge on the base between this and next cell state (Fig.8A). The new cell state (of the same cell or of a daughter cell) will have a new prescription (a new point on a new fiber) and so on; this process will give a tree of cell events in the section s, where each edge connects a point on one fiber with another point on the next fiber.

In the case, when the realization of a prescribed cell event can't be accomplished due to external (injury, cut of the surrounding cells, transplantation, etc.) or internal (molecular/biochemical) constrains (e.g. mutations, DNA damage, chemical or hormonal treatment, etc.), a new possible cell event for a cell in the same *cell state* $\psi$ will be chosen due to a re-organization of local MF according to new situation, taking into account a new set of information in the cell's neighborhood N. This will give an edge connecting an initial point on a fiber to another point inside the same fiber, which we define as an "*Internal cell event*" (Fig 8.B), meaning the changing of a *cell state* $\psi$ without any regular cell event $S_l$.

We will add this type of cell event to the mathematical formalization of Cell events proposed in section 2.2 as following:

Internal cell event: $(A^i_{nm}, \mathbf{D}^i) \rightarrow (Int(A^i_{nm}), \mathbf{D}^i)$.

If the "reorganization" of cell events leads to a realistic new prescription (i.e., to a one which is possible to accomplish in a given molecular state of the cell), a new cell event will be realized and a new edge will appear on the base. If there are several possible realistic new prescriptions, then the rule of minimization of the distance between the initially coded tree of cell events for achieving the final shape of this domain and the practically (biochemically) possible tree of cell events from the current state at a given time point (suggested in statement 2) will be applied.

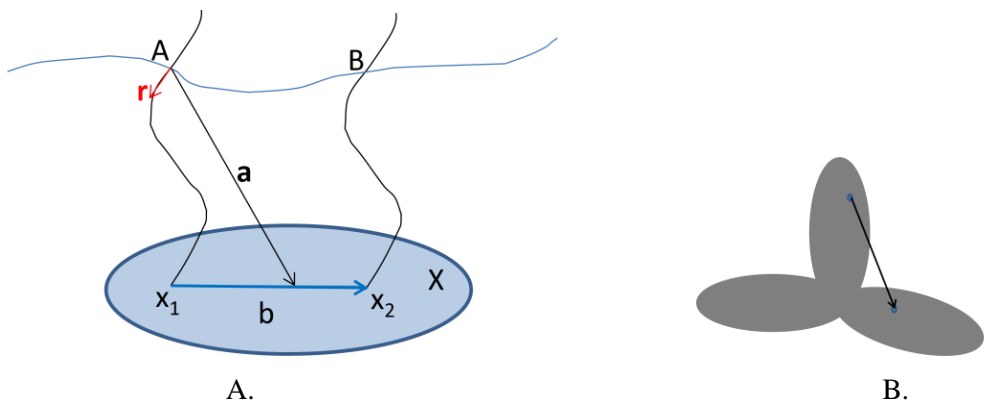

A.                                    B.

**Fig.8 Notion of Realisation of cell event.**
A. Two types of edges, mapping fiber $E_{\tilde{x}}$ (cell events) to a base X or to a fiber itself:
 **a**-as a result of cell event A an edge b on a Base from the point (cell state) $x_1$ to the point $x_2$ appears; **r**- a prescribed cell event A does not occur due to external or internal constrains, and a new cell event is chosen due to re-organization of local MF according to new situation. The edge **r** connects two points inside the same fiber.
 B. "Internal" cell event, which occurs inside the fiber space.

### 4.4 Local MF and a metric space on a base

It is also possible, that the point on the fiber (i.e., the corresponding cell event for a cell) is a function of MF not of the whole "morphological domain" consisting this cell, but rather of a "local" MF, built on the matrices of all cells within the neighborhood N of a particular size around a cell, where the size of N should be determined according to a universal rule. For example, the neighborhood N could be specified as an area on the graph of adjacency corresponding to m number of edges from a given cell (Fig.9); or its size could be calculated using a distance between matrices of cells with a threshold, determining the neighborhood border; or by combination of both factors.

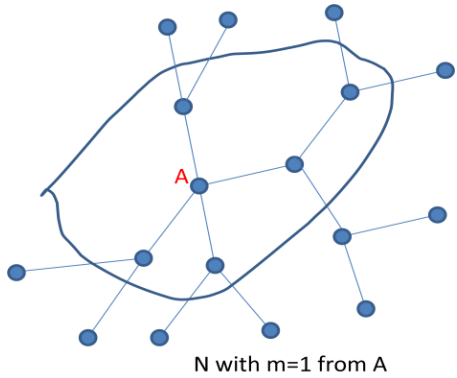

N with m=1 from A

**Fig.9.** Neighborhood N for a cell A on a base, where the size of N is specified as a maximum number of edges of adjacency (m) from the cell A.

We may propose an assumption, that if the general rule for determination of a point in MF by a neighborhood of the size N on the base exists, then the size of N should be different in the case of normal and "crisis" development (e.g., during a regeneration process).

To this end, the question about the "distance" between cells in the embryo is of the utmost importance. One possible proposition could be to use as a "biological measure" for such a distance the distance between "specific cell surface markers of the zygote", which, as was suggested in Section 2.2, may mark the position of cell $i$ in the developing embryo. The existence of these specific zygotic cell surface markers is, in itself, an open biological question.

In general, we may define a metric space on a base- the distance between its points $\psi^i$ (cell states). One possibility to calculate a distance between points $\psi^i$ is to take a sum:

$$d_B(\psi^i, \psi^j) = d(\mathbf{D}^i, \mathbf{D}^j) + d(A^i_{nm}, A^j_{nm}),$$

where, according to cell state definition ( $\psi^i(t) = \bigl(A^i_{nm}(t), D^i(t)\bigr)$, $d(D^i, D^j)$ is the distance between cell positions, $d(A^i_{nm}, A^j_{nm})$ is the distance between cell matrices.

However, calculation of both distances is not a simple issue. For example, several possibilities of defining a notion of cell position in the developing embryo are still under discussion (see section 2.2).On the other hand, the simplest suggestion to calculate a distance $d(D^i, D^j)$ as a distance between cells in R3 at time point t (independently of a notion of cell position itself) may be a plausible approach.

Another helpful suggestion can be to find a the possible correlation between the distance $d(A^i_{nm}, A^j_{nm})$ between cell matrices and the distance between cells measured in edges of graph of genealogy dgenealogy ($\psi^i, \psi^j$).

In any case, the strategy for calculation of the distance between cells (cell states) for determination of the size of neighborhood N should be chosen in such a way, that the resulting distance should be able to discriminate between cells belonging to the same tissue and cells belonging to different tissues, even if the distance in R3 for both cases are the same. The rule for calculation a size of a neighborhood N must take into account only cells from the same domain. For example, in the Fig.10 the distance in R3 between cell C and C' is the same as between cell C and C'', but due to the fact that cell C and C'' belongs to different Tissues (T1 and T2), we may expect that the distance between their matrices should be bigger than between matrices of cells C and C'. Thus, with the proper definition of a distance between cells, e.i., using the matrix information, the neighborhood N for cell C will include cell C', and not cell C''.

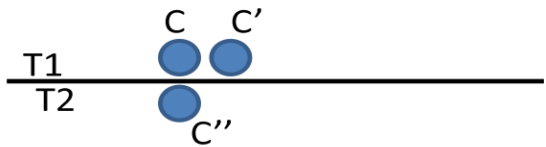

**Fig.10. Question of the notion of a distance between cells.**
    See text for details.

## 5. Probabilistic character of cell events

An additional important suggestion may be to consider next the possibility of a description taking into account a probabilistic character of each cell event $S_l$. The probabilistic character of cell behavior can be seen from the two points:

1. In normal development the final geometry of an organism is coded not unambiguously, but rather as a multitude of possible trajectories within a given defined "corridor", i.e. the form and size of organs may have variations but only within a certain range. This can be represented by a near-constant probability density of cell event paths (MF configurations) inside a corridor, such that the total probability of a path to be inside a corridor is close to unity, with near-0 density outside it.

2. In the case of extreme cell fate changes (via transplantation, placing into cell culture, injury followed by subsequent regeneration, etc.), cells, depending on the level of their totipotency, have the possibility of differentiating into all (or at least several) specialized tissues of the developing organism. This means that at each step of development a cell has many possible alternatives of further behavior, from which only one is realized under given conditions.

Thus, we propose to consider these possibilities as the probability of realization of different cell fate programs. A sequence of cell events that transform an initial state of a tissue $\psi_1$ into a final state $\psi_2$, which we will refer to as a "path", is represented by a particular section of the fiber bundle, $s$, and there is a certain probability that this particular section among the many possible ones, will be realized. Let us denote the probability of transition from $\psi_1$ to $\psi_2$ via a section $s$ by

$$P_s(\psi_1 \to \psi_2).$$

As mentioned above, these probabilities are assumed to have a value that is roughly constant inside the corridor of possible paths, and rapidly falls off outside it. The total probability of reaching a given final state from a given initial one is then given by the sum of individual probabilities for all conceivable paths that start at $\psi_1$ and end at $\psi_2$. That is to say,

$$P = \sum_{s \in \Gamma(E)} P_s(\psi_1 \to \psi_2)$$

where $P = P(F_c[s])$, $\Gamma(E)$ is the set of all possible sections s of a bundle space $E$. Clearly, the probability of a particular path depends on the probabilities of individual cell events comprising it, in the usual way:

$$P_s(\psi_1 \to \psi_2) = \prod_i P_i(S_l)$$

where $P_i(S_l)$ is the probability of an individual cell event occurring. However, since not every path constructed in such a way is possible, e.g. a cell that has divided at a previous moment in time cannot perform growth in the next (simply because it no longer exists), the cell events $S_l$ in the above product must all belong to the same section, connecting the initial and final states.

Therefore, it would be advantageous, if possible, to find a way to characterize the probabilities of entire paths, rather than having to build them up from individual events. For this, we propose an idea to account for the probabilistic nature of cell events using a formalism where the amplitude for a given path is a function of the action evaluated along that path. In this case, there exists a "classical path" of events (corresponding to the initially coded pathways of cell events), and all the paths in its vicinity contribute the most to the probability of transitioning from the initial state to the final state. Paths that lie further away from the "classical" one also contribute, but with quickly declining weights. Thus, by taking the sum of $P(F_c[s])$ on $\Gamma(E)$, where $\Gamma(E)$ is the set of all possible sections of $E$, one will get the situation where the main contribution to probability will be given by the sections that satisfy the equations following from (7,8), but other possible trajectories of cell motion will also potentially exist. Finding such a characterization is a work in progress.

**6. Evolution of MF as a tree of hypersurfaces**

If the MF of a developing organism consists of many local sections in the total space of the bundle, corresponding to all morphological domains, then there exist species-specific surfaces in this space, determining a branching of the initial MF section to new ones, corresponding to the appearance of new MFs of new domains, tissues and organs. These surfaces correspond to the border between cell events Differentiation (and maybe also Apoptosis), and cell events of other types (Fig.11).

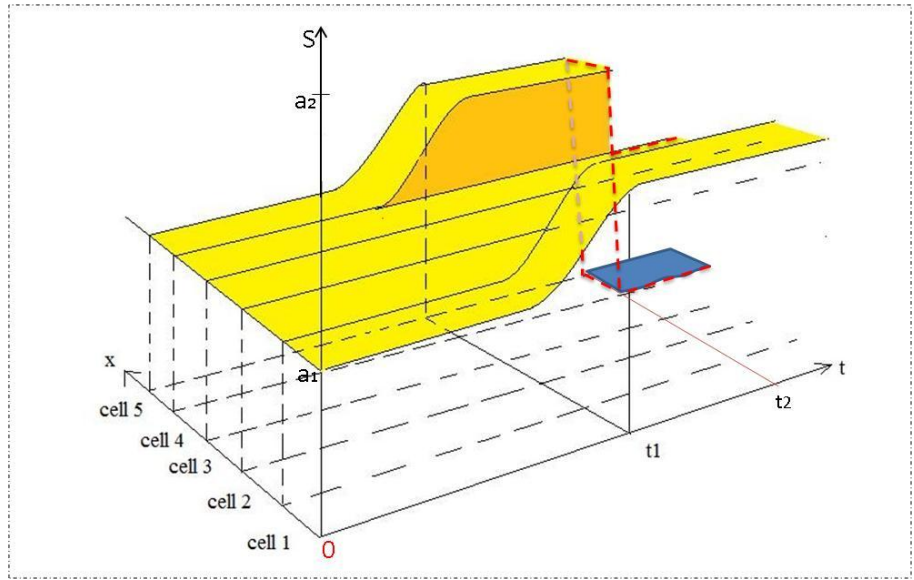

**Fig. 11. A simplified model of MF in a fiber bundle representation.**
Five cells are located in 2-dimentional space (x,t), and correspond to the same section of MF above them (colored in yellow). Until the point in time $t_1$, all five cells are performing the same event (e.g. growth in the same direction with the same growth rate $a_1$). At time $t_1$, cell 1 and cell 5 start to grow with different rate, and the section over them obtains a new value on the fibers S ($a_2$). At time $t_2$, cell 1 still continues to grow, thus the section above it is still the same one (yellow), while cell 5 starts to perform an event, differentiation, which means the appearance of a new section (in blue). The borders between different sections are shown by a red dotted line.

It is possible that, for each living species, not only the final geometrical form is determined, but also the set and sequence of appearance of local MFs, corresponding to each morphologically distinct domain of the embryo, that is the plan of boundary surfaces in MF space corresponding to species development is determined. It can be understood as a "tree of surfaces". When the MF of a tissue, at some point, reaches a determined surface in total space, the cell on its border undergoes one of the following processes: non-symmetric division (meaning that two daughter cells will be committed to two different MFs (two different tissues), death, or final determination. New daughter cells after non-symmetric division "on" this surface will belong to two different new local fields, or one to a new and one to an "old" field.

This means that, if we present development as a tree of cell events, then **the structure of this tree has a freedom/flexibility inside each determined local surface, belonging to different MFs, and this freedom is stipulated, among other factors, by a probabilistic character of cell events.** At the same time, there exists a law determining the branching and/or separation of these surfaces in MF space.

## 7. Determination of $R_l$, $C$ and $M_F$

The next step in formalization is to discover equations/rules/laws determining $F_c$, $C$, $R_l$ and $M_F$, associated with a given biological information of the object.

We assume that the general form of the functions $C$, $M_F$, and $R_l$ should be the same for all organisms, while its realization for each species should depend on the concrete biological information of an organism (on the corresponding matrices of the oligosaccharide code). We propose that determination of equations for the function $C$ may be possible by using the understanding of a morphogenetic field as a section of a fiber bundle over the cell space. The general form of the function $C$ may be inferred by comparison of experimental data from early steps of embryogenesis for different types of organisms (e.g. sea urchin or dicotyledonous plant (Fig.12 a, b)), with the idea of finding common general features between them using cell events formalization.

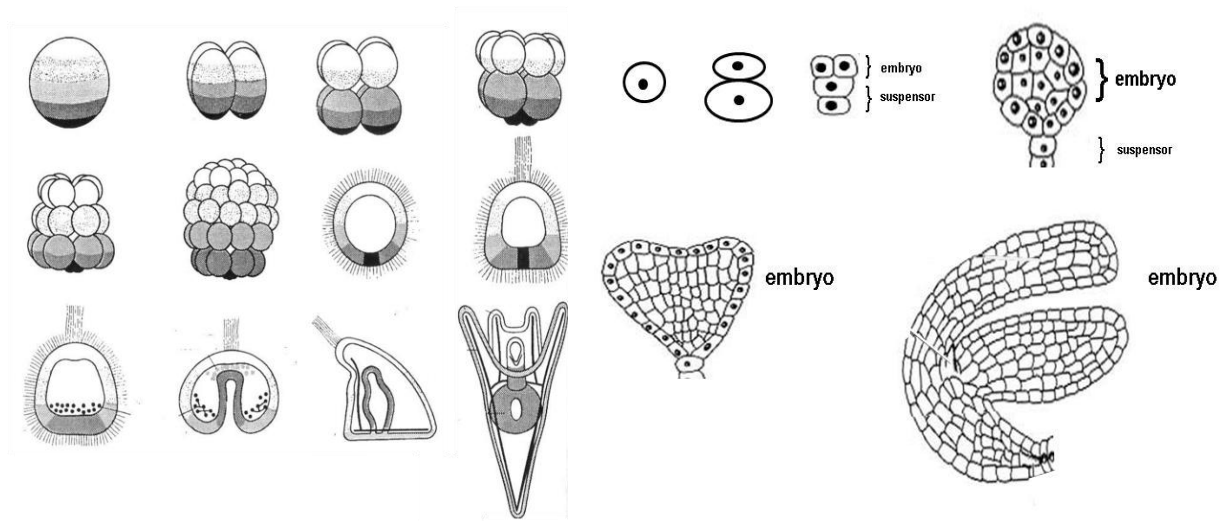

A.          B.

**Fig.12.A. Schematic representation of the initial steps of sea urchin development, starting from one cell (zygote) (after Horstadius, 1939). B.Schematic representation of the initial steps of dicotyledonous plant development, starting from one cell (zygote).**

The main open question for creating a real formalization of pattern formation in the framework of the suggested model is the one regarding interrelation between functions $F_c$, $C$ and $M_F$. Our working hypothesis is that this interrelation could be found mathematically using the language of differential geometry (e.g. the analogs of the notions of connection, etc.) and the analogs of the notion of potential, adapted for the proposed model. However, this work is only in the preliminary stages and strongly depends on all other details of the model.

After getting the general equation for $M_F$, it might be possible to find the equation for the corresponding cell event $S_l$:

$$S_l = S_l(M_F)$$

On the other hand, the correspondence between cell event $S_l$ and the rule $R_l$ for changing the oligosaccharide code (matrix $A^i_{nm}$) on the cell surface might be found experimentally. All these steps may give the final result – the law of creation of a corresponding MF (for example, its $M_F$) by a set of $A^i_{nm}$ of corresponding cells.

An additional question, which may also be an important and non-trivial one, is about the interrelation between the shape of an embryo domain, described here as an equation for its surface *G(t)*, and the suggested MF formalism. The current model proposes that the influence of the MF laws on the evolution of an embryo's domain surface occurs via an intermediate "cell events" step. However, it is possible to suggest that the vector function $\sigma(x,t)\mathbf{n}(x,t)$ has a potential, in which case the family of surfaces *G(t)* would be described by the contours of this potential. On the other hand, it is possible to suggest that a function $M_F$, corresponding to MF influence, and a cell motion function $C$ might be expressed in terms of the potential of MF. In this case we may have a question: is it possible to find a direct interconnection between these two, generally speaking, differently originated potentials? For example, the bases for such interconnection might be the following.

When viewed from a more intuitively tangible perspective, the effect of a morphogenetic field is to cause growth and development of cells at a given location. Thus, it corresponds to a change in form. The potential of such a field, if it exists, would therefore correspond to form itself (where by "form" one may understand both geometrical shape and the type of tissue comprising it). That is to say, that a potential, which is time-dependent, would represent the form that an organism tends to achieve at a particular time. This is in close parallel with the idea of the potential of the vector field of the rate of growth, whose contours would also correspond to the shape of the embryo, or domain surface at particular times. The mathematical verification of this parallel is another question for consideration.

Although rather loosely defined, such a potential satisfies the requirement of being independent of path: under most ordinary conditions, the final form of an organism is not dependent on the path that it took to arrive at it. This is because, in the above interpretation, the potential is associated with form itself, rather than with an analog of physical work. Indeed, the "work" (an analog of work, defined in terms of cell events or otherwise) done to achieve one and the same form through different paths may be significantly different. A rather obvious example would be the process of limb formation in the case of normal development and in the case of full regeneration of the whole limb from a small segment of limb disc, where different "work" will be done on the same set of cells in the segment in order to achieve the proper form of the limb. Meanwhile, the same final shape is achieved through two different paths. However, it is clear that the appropriateness of this analog is an open question, important for further development of the proposed formalization.

### 8. Suggestions for biological and numerical experiments

It is very important to note that, with or without considering the suggested MF theory, the work presented here contains several assumptions that would be interesting to test experimentally and numerically.

The first problem which we would like to suggest for experimental study is the assumption of the existence of an oligosaccharide code on the cell surface, and next, if this is shown to be the case, to determine the most meaningful way of presenting this information in matrix (or other formalized) form. It is possible that a biological code on a cell surface similar to that proposed can be found, but in another biological form (i.e. not as oligosaccharide residues). The experiments, monitoring the changing of the different types of cell surface cell markers in the course of different cell events, could be very promising.

The second experimental problem is to find the correspondence between a cell event $S_l$ and the rule $R$ for changing the biological code (matrix) on the cell surface. Both these tasks require experimental research and mathematical work (matrix analysis) to be carried out in tight collaboration.

The third problem is to give a mathematical description of the dynamics of the boundaries (surfaces) of the morphological domains of an organism in a space-time as a function of the set of all cell events that have occurred inside the volume enclosed by this surface during a time period $dt$. This investigation can be undertaken in a computational experiment based on the detailed descriptions of the early stages of embryogenesis for several types of organism, taken from the published data. To this end, the equation of evolution of a domain/embryo surface will be an important step. The equation for the smooth evolution of a surface, corresponding to $\sigma(x, t)$ approximated by equation (2), generally cannot be solved by regular methods. However, with some simplifications, regular methods might give good results (Courant, Hilbert, 1962). Furthermore, as differential equations deal with continuous functions, and cell events are discrete, "difference equations" may be a possibility.

### Conclusions:

In this work we suggest several important assumptions/notations for discovering the interconnection between biological information and the geometrical form of an organism:

1. The geometry of an organism and its parts is coded by a biological code located on the cell surfaces. As a particular model, we suggest a code of 6 types of oligosaccharide residues of glycoconjugates.
2. For each cell this code can be presented as a corresponding matrix.
3. The notations of *cell state* and *cell event* are given.
4. The matrix, corresponding to a cell, is changed after each cell event (cell division, growth, etc.) according to a specific rule, $R_l$, for which we have suggestions for decoding.
5. The geometry of an organism during its growth and development can be described in terms of dynamics of the boundaries (surfaces) of the morphological domains of an organism.
6. The dynamics of the surfaces of the domain depends on the set of all cell events that have occurred inside the domain enclosed by this surface.
7. The notion of morphogenetic field is given as an object governing a law(s) for decoding the biological (cell surface markers) information and transforming it into corresponding instructions for cell event depending on the positional information (location) of a cell in the developing organism.
8. A mathematical formalization of morphogenetic field is suggested.
9. A set of biological experiments for investigating the proposed theoretical hypothesis is suggested.

**Acknowledgements:** The authors are very grateful to Christophe Soule and Pasha Zusmanovich for very fruitful and essential discussions of the second version of the manuscript.